# *New measurements of neutron electric dipole moment with double chamber EDM spectrometer*

A.P.Serebrov[1], E.A.Kolomenskiy, A.N.Pirozhkov, I.A.Krasnoshekova, A.V.Vasiliev, A.O.Polyushkin, M.S.Lasakov, A.N.Murashkin, V.A. Solovey, A.K.Fomin, I.V.Shoka, O.M.Zherebtsov

*B.P.Konstantiov Petersburg Nuclear Physics Institute of National Research Centre "Kurchatov Institute", 188300 Gatchina, Leningrad region, Russia*

P.Geltenbort, S.N.Ivanov, O.Zimmer

*Institut Max von Laue – Paul Langevin, BP 156, 38042 Grenoble Cedex 9, France*

E.B.Alexandrov, S.P.Dmitriev, N.A.Dovator

*Ioffe Physical Technical Institute RAS, 194021 St. Petersburg, Russia*

**Abstract**

The article presents results on neutron electric dipole moment measurements (EDM), made by ILL reactor using PNPI experimental installation. Double chamber magnetic resonance spectrometer with prolonged holding of ultra cold neutrons has been employed. The obtained results at 90% confidence level determine the upper limit for EDM neutron quantity equal to $|d_n| < 5.5 \cdot 10^{-26}$ e·cm.

## *1. Introduction*

The issue of symmetry is one of the key concepts in modern physics. Invariance of elementary process relevant to symmetry transformation implies a certain character of physical laws. One of the fundamental theorems of the quantum field theory is CPT-theorem, according to which all physical interactions are invariant with respect to combined CPT-transformation, where C, P and T are discrete symmetry transformations: operation of charge conjugation, space and time reversal respectively. Elementary particle can possess electric dipole moment (EDM) only in case of simultaneous violation of P- and T-symmetry.

As far back as 1950 Purcell and Ramsey were the first to point out the necessity of experimental testing the possibility of elementary particles to possess EDM [1]. A few years later there appeared information on the first experiments on search for neutron EDM. Interest in this problem has considerably grown after discovery of P-parity violation in weak interaction [2,3] and especially after experimental finding of CP-parity non-conservation in decays of neutral K-mesons [4]. In the macro world an indirect evidence for processes occurring with violation of CP-parity is asymmetry of matter and antimatter observed in our Universe. Understanding of the character of fundamental symmetry violation is supposed to throw light on the problem of origin and evolution of Universe at the very first stages.

The standard model of electro-weak interaction (SM) provides estimations on neutron EDM value on the level inaccessible for the modern experiment: $10^{-30} - 10^{-33}$ *e*·cm. CP-violation (and neutron EDM) arises here only in the second order of smallness on the weak interaction constant. Hence, SM fails to definitely account for baryon asymmetry of Universe. Thus, search for neutron EDM is expected to be search for some phenomena beyond the framework of SM.

Gauge theories on spontaneous symmetry violation gave rise to creating models of another kind for describing CP-non-conservation [5, 6]. Such models, as super symmetry ones, those with numerous Higgs particles and left-right symmetry theories involve extension of SM symmetry and add new particles. EDM value in these models arises in the first order on weak interaction and turns out to be equal to $10^{-26} - 10^{-28}$ $e\cdot$cm. Finding neutron EDM at such a level or a new constraint on its value is essential for choosing the theory adequately describing phenomena of CP-violation.

---

[1] E-mail: serebrov@pnpi.spb.ru

Particle electric dipole moments are regarded as being sensitive probe for such new physics, complementary to search for new particles by high energy accelerators. While the latest measurement on the electron EDM [7] seems to challenge the minimal super-symmetric model of electroweak baryon genesis [8], further sensitivity gain in search for particle EDM and notably that of the neutron will be needed for a final discovery of the required additional CP violation [9].

To illustrate it in coordinates of mass of particles being sought for in super symmetric theoretical models, Fig.1 shows areas (for possible existence of super particles) excluded by other experiments [8].

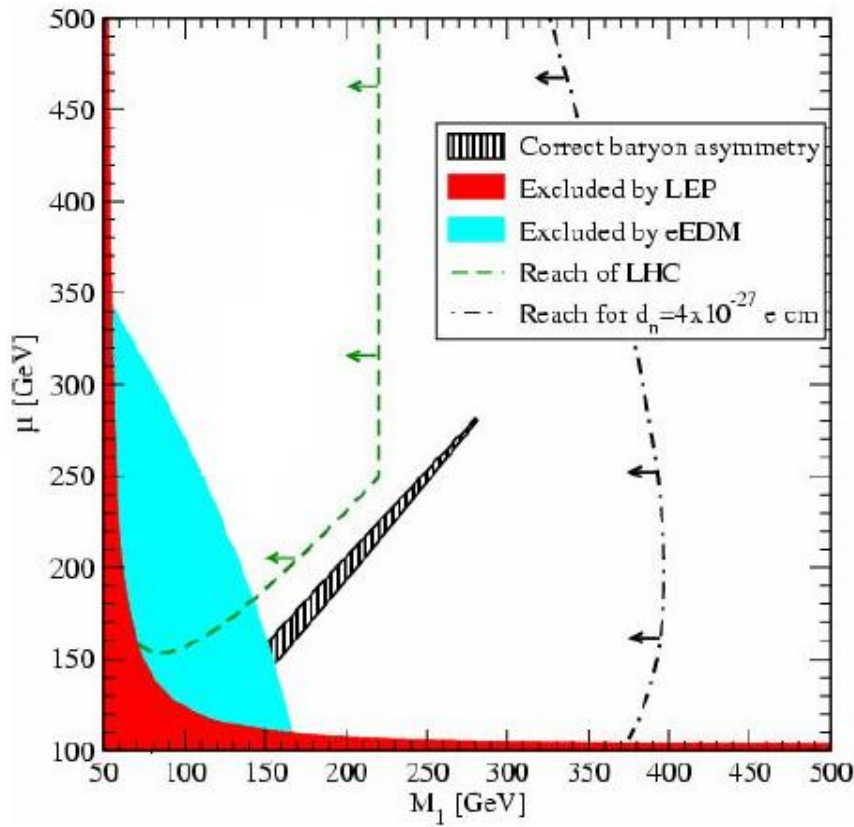

Fig. 1. Exemplary sensitivity of present and future EDM measurements
on super symmetric mass parameters relevant to electroweak baryon genesis [7, 8]

The vertical axis shows the super symmetric Higgs/Higgsino mass, while the horizontal axis gives the U(1)Y gaugino mass parameter M1. The shaded region presents the region for the masses to obtain the observed baryon asymmetry. The dark grey region is excluded by LEP. The grey region is excluded by the present electron EDM limit, while the dash-doted line gives experimental sensitivity of neutron EDM to be attained. The dashed line indicates sensitivity of further direct search measurements at the LHC. The area accessible to the Large Hadrons Collider (LHC, Switzerland) is shown with a dotted line. The area accessible for EDM experiment in increasing the experimental accuracy by two orders of magnitude is marked with a dot-dashed line. It will be quite possible to achieve it using new UCN sources of high intensity.

For experimental search for neutron EDM one employed both the magnet resonance method suggested by Ramsey at polarized neutron beam [10, 11] and the neutron diffraction method at crystals [12] which investigated interaction of neutron spin with electric field inside the crystal. The main difficulty in experiments on beams of thermal and cold neutrons is related to systematic effects imitating the searched effect produced by neutron EDM. First of all, it is concerned with the effect caused by Schwinger interaction of magnetic moment of neutron moving in the electric field. Hence, accuracy obtained in investigations performed earlier was limited to a systematic error rather than a statistical one, and experiments of such kind provided constraint on the upper limit for neutron EDM existence at the level $|d_n| < 3 \cdot 10^{-24}$ $e \cdot$cm [11].

PNPI scientists suggested using non-center symmetrical crystal for a crystal diffraction experiment providing efficient electric field at a very high intensity for the neutron to be influenced. This condition as well as sufficiently long time for the neutron to pass through the crystal Bregg angles close to $\pi/2$ can ensure sensitivity of such an experiment comparable with UCN method [13, 14]. The first experiments have been already carried out [15].

At present UCN application for EDM experiment with neutron storage appears to be the most promising direction to concentrate on. UCN kinetic energy is less than Fermi pseudo potential of some materials, due to which neutrons under complete internal reflection from the walls of the storage volume can remain in closed vessels for a long time [16, 17]. It enables to increase considerably experimental precision by enhancing the time of neutron interaction with the device. The main EDM imitating a systematic false effect caused by swinger interaction of neutron magnetic moment with electric field, in this case, is much suppressed because of lack of evident direction of UCN movement during their storage. Sensitivity of our installation and final experimental precision were mainly determined by the number of registered neutrons. At considerable enhancement of UCN source intensity, installation sensitivity will be improved, hence, perspectives of developing EDM experiments depend on creation of a novel generation of UCN sources.

Investigations on search for neutron EDM by UCN method were first performed in 1980 in PNPI (Gatchina, Russia) [18, 19] and then in ILL (Grenoble, France) [20, 21, 24]. The first constraint on neutron EDM obtained by UCN approach comprised $|d_n| < 1.6 \cdot 10^{-24}$ $e \cdot$cm (90% C.L.) As early as 1981 the result obtained in Gatchina was improved: $|d_n| < 6 \cdot 10^{-25}$ $e \cdot$cm (90% C.L.) [19]. In the 1990-1991s both research teams attained EDM limit $\sim 1 \cdot 10^{-25}$ $e \cdot$cm (90% C.L.) [21, 22, 23]. At this stage the measurements made in Gatchina were stopped because the UCN source ceased working. In Grenoble measurements were continued by collaboration of RAL/Sussex/ILL, and 15 years later the neutron EDM limit became 3 times lower [25]. The investigation concerned had the best ever obtained constraint on the value of neutron EDM $|d_n| < 2.9 \cdot 10^{-26}$ $e \cdot$cm (90% C.L.). One used in it a chamber for storing UCN and a mercury co-magnetometer for monitoring magnetic conditions. Unfortunately, the mercury magnetometer cannot be an absolute co- magnetometer because of the so - called geometric phase [26], varying for neutrons and mercury atoms. It is responsible for a systematic error in case of gradient of the magnetic field.

The present work result is somewhat worse than that attained in work [25], however, it was obtained by employing an experimental installation based on another approach. We apply a differential magneto-resonance spectrometer with a double chamber for storing UCN, with common constant magnetic field and with electric fields of opposite directions in operational volumes for neutron storage. It provides a principally different possibility for controlling systematic effects. In the course of making measurements at the attained accuracy level no systematic errors have been found.

Fig. 2 shows chronology of decreasing the upper neutron EDM limit in experiments conducted in Gatchina and Grenoble.

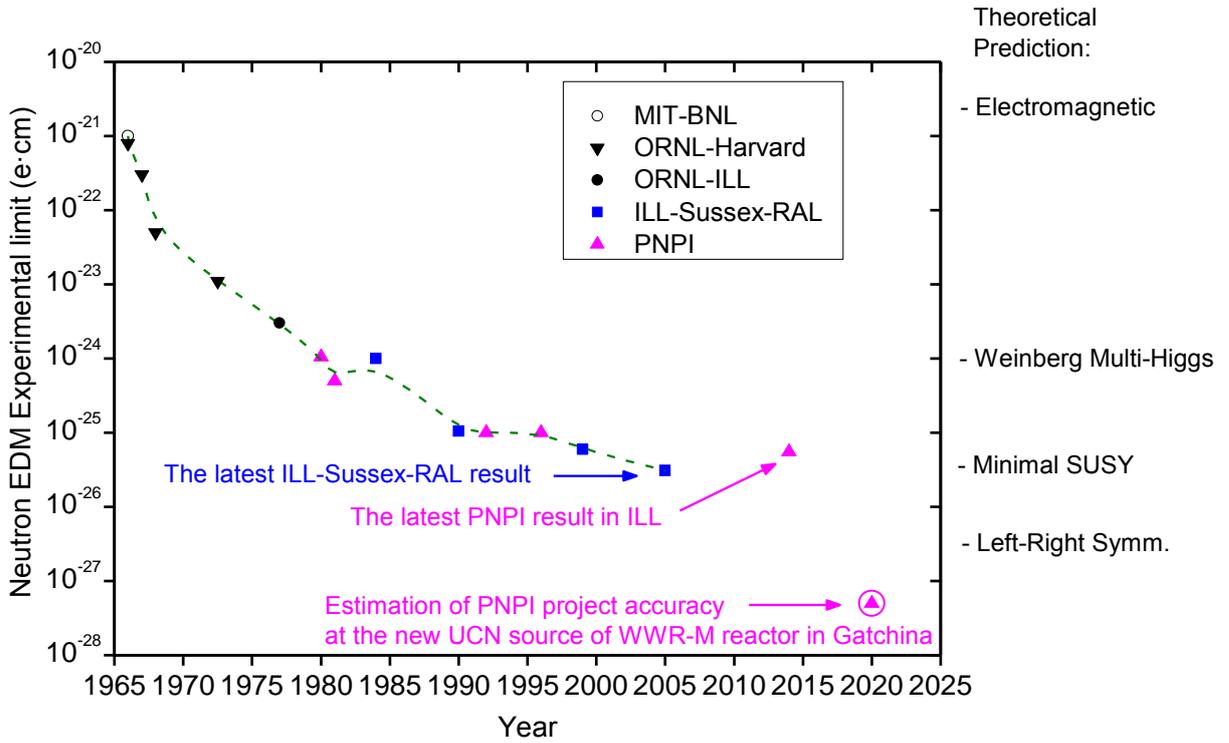

Fig.2. Stages of decreasing an experimental limit on neutron EDM and perspectives for precision increase.

The tasks of EDM-experiment became a crucial impetus for developing a novel technique of ultra cold neutron production such as UCN cryogenic sources, located either in close vicinity to or in the reactor core itself [27, 28]. However, UCN sources available now do not allow to expect much better result than that obtained. Nowadays in a few foreign scientific centers studies are being in progress on elaborating new ultra cold neutron sources: ILL (France), LANL (USA), PSI (Switzerland), TUM (Germany). PNPI (Gatchina, Russia) is planning to create UCN sources of high intensity based on superfluid He, with the reactor PIK under construction and with the operating PNPI reactor WWR-M [29, 30]. The calculated UCN density of these sources is by 2-3 orders higher than that in the available source of ILL. Such sources enable to obtain EDM estimation accuracy at a higher level than $10^{-27}$ $e$·cm, as shown in Fig.2. This level of experimental precision will make it possible to estimate prediction validity of different theories on CP-violation (super-symmetric, left-right ones and others).

The present work describes the double chamber differential magneto-resonance EDM spectrometer made in PNPI and used in the beam of PF2 MAM of the reactor ILL (Grenoble, France), with the latest measurement results quoted.

2. *The method of measuring.*

The direction of neutron spin is the only distinguished direction in the system of coordinates related to neutron, thus EDM vector $\mathbf{d_n}$, if it exists, must be collinear to the neutron spin. In the external magnetic field $\mathbf{B_o}$ potential energy of magnetic moment interaction $\mathbf{\mu}$ with the field W= −($\mathbf{\mu}·\mathbf{B_o}$) and the neutron energy state splits into two levels corresponding to two possible projections of the neutron spin into the field direction.

If within the magnetic field a constant electric field collinear to $\mathbf{E}$ is applied , while neutron EDM is different from zero, then potential energy of neutron interaction will be expressed by the following equation:

$$W = -(\mathbf{\mu}·\mathbf{B_o}) - (\mathbf{d_n}·\mathbf{E}).$$

Due to an extremely low value of $d_n$, additional to the neutron interaction energy is very little, yet it must result in frequency shift of magnetic resonance in the available electric field. If $\mathbf{E}$ and $\mathbf{B_o}$ are parallel and dipole moment is $d_n > 0$, then magnetic resonance frequency decreases:

$$\omega = 2\mu \cdot B_o \cdot \hbar^{-1} - 2d_n \cdot E^+ \cdot \hbar^{-1}.$$

Fig.3 gives the scheme of neutron energy states under the mutual influence of magnetic and electric fields for cases when electric field is parallel and anti-parallel to the magnetic field.

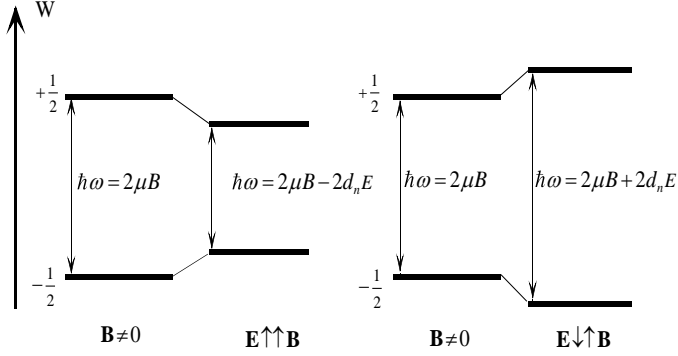

Fig.3. Scheme of neutron energy states in magnetic and electric fields.

Reversal of the electric field direction causes frequency alteration of spin precession: $\Delta\omega = 2d_n \cdot (E^+ + E^-) \cdot \hbar^{-1}$. Hence:

$$d_n = \frac{\Delta f \cdot h}{2 \cdot (E^+ + E^-)},$$

where $E^+$ and $E^-$ – electric field intensities when **E** is directed parallel and anti-parallel to $B_o$; $\Delta f = \Delta\omega/2\pi$ – frequency shift of neutron spin precession; $h = 2\pi\hbar$ – Plank constant.

Ramsey method of discrete oscillating fields is the most sensitive one for recording frequency alteration of spin precession [31]. In our case of using a spectrometer with UCN storage at the beginning and at the end of neutron storage, one applies two coherent pulses of oscillating magnetic field **B₁** with frequency $f_o = 2\mu \cdot B_o h^{-1}$. The first pulse directs spin of polarized neutrons into the plane perpendicular to the leading magnetic field. During the storage time $T_s$ neutron spins are involved in coherent precession around $B_o$, if the required homogeneity of constant magnetic field is provided. If a phase of the second radio frequency pulse coincides with that of the first one, then neutron spins will be additionally turned at 90°, thus having changed the original polarization to the opposite one. At 180° phase shift the original polarization will be restored. If the phase shift between oscillating fields is equal to 90° or 270°, then neutron spins will remain in the plane perpendicular to $B_o$. In this case their mean polarization at $B_o$ direction will be equal to zero, unless additional phase alteration of neutron spin precession occurs (for example, due to dipole moment interaction with electric field). Neutron intensity recorded at the spectrometer output behind the polarization analyzer depends on the oscillating field frequency.

Fig.4 presents resonance curves of count dependence of two installation detectors on linear frequency of the oscillating field during neutron storage $T_s = 95c$ and phase shift between two radio frequency pulses 90° and 270°. Frequency scanning was performed by a special electronic scheme with a controlled division fraction coefficient, which produced the required oscillating field frequency from the mean frequency of eight cesium magnetometers in correspondence with the mean magnetic field influencing neutrons in the UCN storage volume.

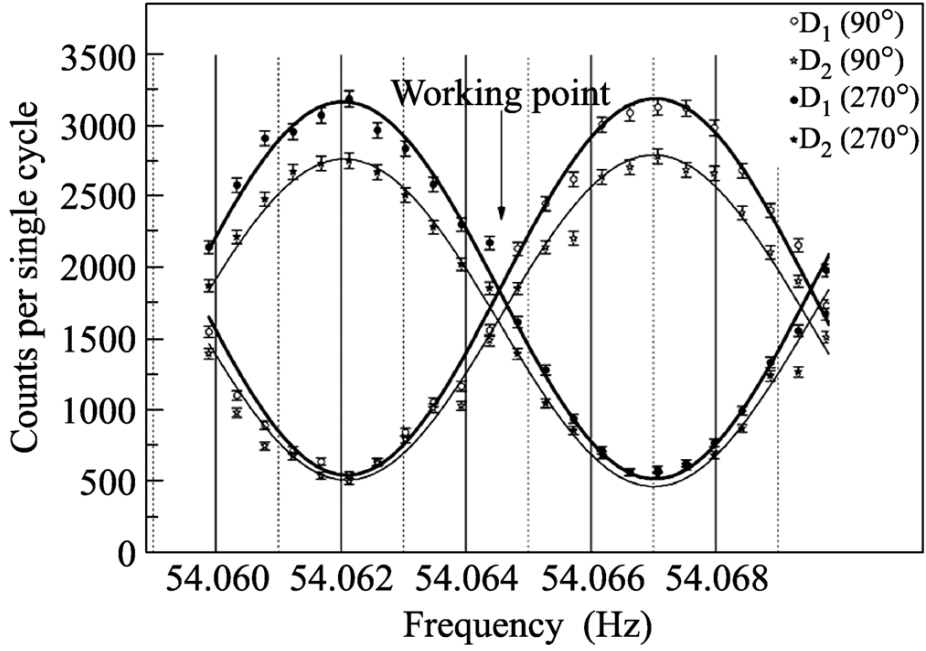

Fig. 4. Central parts of Ramsey resonance curves for detectors D1 and D2 and phase shifts 90° and 270° between rf pulses. The visibility of the resonance was 0.7 for $T = 95$ s.

Thus, alteration of the detector count in changing precession frequency of neutron spin is the highest at the resonance point at the intersection of curves with a phase shift between separate oscillating fields 90° and 270°. The shape of the resonance curve in the vicinity of the working point is described by sinusoid:

$$N(f) = \alpha \cdot N^{res} \cdot \mathrm{Sin}(2\pi\, T( f - f_o)) + N^{res},$$

where $\alpha = ( N_{max} - N_{min} )/( N_{max} + N_{min} )$ – parameter characterizing the resonance curve range; T – storage neutron time; $N^{res}$ – neutron count at resonance frequency.

The longer is the time of neutron storage in the resonance area, the narrower is the resonance curve and the higher the installation sensitivity to frequency alteration of neutron precession. Difference in detector counts after two periods of storage at the opposite directions of electric field concerned with frequency shift of magnetic resonance and slope of resonance curve at the working point is expressed by the following ratio:

$$( N^+ - N^- ) = \Delta f \cdot ( \partial N/\partial f ) = \Delta f \cdot 2\pi\alpha\, N^{res}\, T.$$

The derivative sign depends on choosing the phase between two radio frequency pulses. Thus, one gets the following equation for neutron EDM:

$$d_n = \frac{h(N^+ - N^-)}{2(E^+ + E^-)\frac{\partial N}{\partial f}}, \qquad (1)$$

where $N^+$ and $N^-$ – detector counts at parallel and antiparallel directions of the electric field related to $\mathbf{B}_o$, correspondingly, $( \partial N/\partial f )$ is the resonance curve slope at the working point.

Statistical error of the measured EDM is expressed with the following formula:

$$\delta d_\mathrm{n} = \frac{h\sqrt{2N_{res}}}{2(E^+ + E^-)\cdot \frac{\partial N}{\partial f}} = \frac{h\sqrt{2N_{res}}}{2(E^+ + E^-)\cdot \alpha \cdot 2\pi \cdot T \cdot N_{res}} \ , \qquad (2)$$

where $2N^{res}=(N^+ + N^-)$ – total neutron count at resonance frequency, accumulated during measurements, T – time interval between two pulses of oscillating field.

Summarizing, experimental sensitivity is determined by neutron count $N^{res}$, interaction time in the installation T and value of electric field intensity E. Range of the resonance curve (derivative at the resonance point) depends on both initial polarization of neutrons and on condition of their storage in the trap. Inhomogeneity of the magnetic field in the storage volume and collisions with the trap walls result in partial neutron depolarization, thus decreasing the installation sensitivity.

### 3. Differential double-chamber EDM spectrometer.

Double-chamber EDM spectrometer with reversible electric field for EDM neutron estimations was described in the article [23]. The spectrometer scheme and the installation outlay are shown, respectively, in Fig.5 and 6.

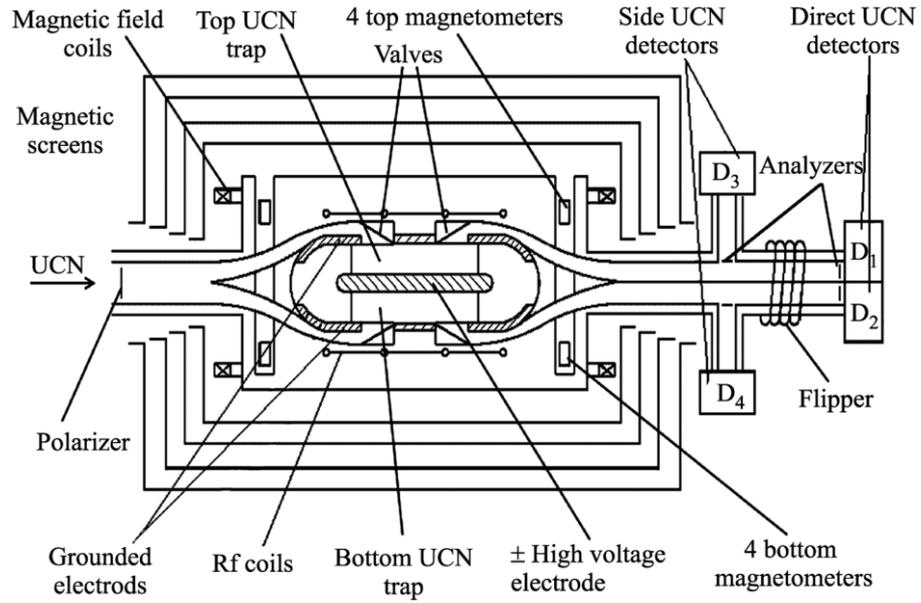

**Fig. 5.** Schematic side view (cut through the center) of the EDM spectrometer.

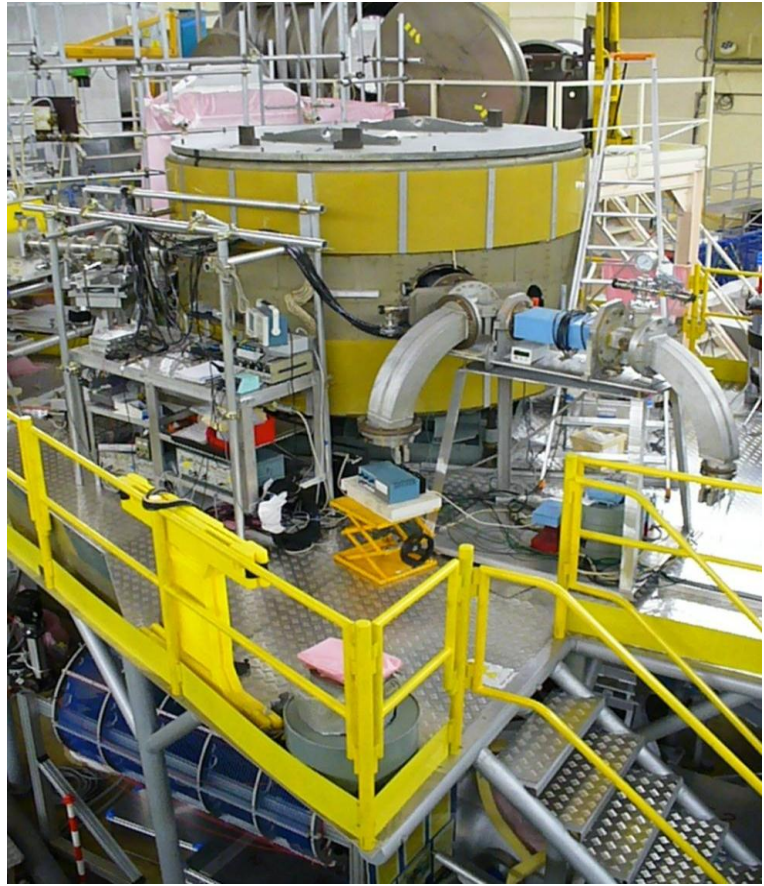

Fig. 6. General view of the PNPI spectrometer in the experimental hall of ILL.

Specified feature of our spectrometer is that it has two chambers for UCN storage with the mutual system of magnetic fields and electric ones equal in value but of opposite directions. In changing polarity of the electric field, neutron EDM effects in different chambers will have opposite signs, while instability of common electromagnetic conditions gives rise to the shift of resonance frequency of the same sign. Difference in results of these estimations leads to additional effects produced by neutron EDM, with the effects from correlated count alteration irrelevant to EDM being greatly suppressed.

Another distinctive characteristic, enhancing the installation sensitivity is the system of double polarization analysis. At the output from every chamber there are two detectors, each of them recording the definite neutron polarization component with respect to the leading field. As the direction of the leading field in the output neutron guides is preserved, spin flipper is placed in front of the secondary detectors for registering the secondary polarization component. This increases summary neutron count during estimations and allows to compensate the data spread resulting from intensity fluctuations of neutron sources. Data analysis from four detectors enables to find out systematic effects.

It should be noted that the four detector system does not have equally high aperture. As a rule, UCN count in the top chamber is higher than that in the bottom one. Also is different the UCN count for detectors located in front of flipper and behind it. These differences are accounted for by various gravitational location of storage chambers and various geometry of location of detectors and polarization analyzers – either onward the UCN beam or to the side direction from that. Such a symmetry in UCN count on different detectors somewhat reduces compensation mechanism to systematic errors of the spectrometer scheme with double chambers and the system of double polarization analysis.

**The main installation units and parameters are to be given in brief.**

*Neutron guides*. Ultra cold neutrons are transported to the polarizer by a rectangular neutron guide with cross-section 70×60 mm$^2$ made from stainless steel, with operational surface sprayed with isotope alloy $^{58}$Ni(90%)Mo(10%) at boundary velocity for UCN about 7,8 m/s. The neutron guide area behind the polarizer is made from non-magnetic materials such as metallic ones (copper), glass replica and high quality surface glass and has nickel molybdenum alloy spraying. This neutron guide is divided into two parts and transports neutrons via covers-electrodes and the top and bottom UCN storage chambers.

*Polarizer*. Polarization and the neutron polarization analysis are performed in the input and output neutron guides respectively, beyond magnetic screens by passing UCN through alloy films Fe(55%)Co(45%) of 200 nm thickness on the titanium substrate as thick as 120 nm and magnetized to saturation in the field of constant magnets. Neutron polarization after passing through the polarizer film was ~90%.

*Neutron detectors.* Proportional 3He-counters with aluminum ports as thick as 0,1мм were used as detectors. To reduce neutron loss on aluminum foils in front of detectors, neutrons were accelerated in gravitational field with Δh=0,6m. Efficient neutron registration is 80%.

*UCN storage chambers.* UCN storage chambers are located in the aluminum vacuum cell surrounded by magnetic screens. All spectrometer units inside magnetic shielding are made from non-magnetic materials such as aluminum, titanium, beryllium bronze, fluorine layer or glass. Operational vacuum in the spectrometer ~10$^{-5}$ mbar was created by turbo- molecular pumps and maintained during a long period of time with cryo-sorption pumps.

Lateral walls of UCN storage chambers are formed by rings made from isolating material with high volume electric resistance. The height of rings is about 100 mm, with inner diameter being 526 mm. Top and bottom storage chambers are separated by a mutual aluminum electrode with beryllium coating at high voltage. Top and bottom covers of neutron storage chambers are also made from aluminum with beryllium coating and are always under zero potential of high voltage source. Electrical contacts of current input conductors with electrodes are connected in the electrode center. This provides a symmetrical distribution of current leakage in neutron storage chambers and reduces the vertical component of magnetic currents.

Melted quartz and sitall were used for rings-insulators. Sitall was proposed to be used for EDM experiment in work [32]. Sitall is microcrystal material characterized by high solidity, mechanical firmness, thermal and chemical stability and perfect electro- isolating properties. Sitall does not practically absorb water, its structure has neither pores, nor rough capacity defects or inhomogeneity, thus ensuring high electric stability. Bulk resistivity of our sitall is ρ>10$^{13}$ Ohm·m (20°C). Specific feature of sitall is that one can obtain compounds having both negative, zero and high positive coefficient of thermal expansion. It allows receiving conjugates on thermal characteristics consistent with other materials including metals.

It is shown experimentally that preliminary heating of chambers for neutron holding up to ~200°C in vacuum or helium atmosphere decreases time of high voltage to a great extent, reduces leakage currents and enhances UCN storage time. The present installation does not permit to carry on heating of storage chambers in assembly and directly in the spectrometer vacuum chamber itself. This was done periodically in a special box with a subsequent transfer of some units into the spectrometer via air, which undoubtedly diminished effect of heating. It is desirable that heating of storage chambers should be performed in assembly and directly in the spectrometer vacuum chamber itself.

In this case, beside the problems concerned with choosing heaters and thermo-isolation for the other part of the installation, there arises the issue of finding suitable materials for high voltage electrodes and cylinder isolators. In order to prolong the time of UCN storage, gaps between electrodes and isolators must be minimized and the latter should be made from materials with very similar (or better the same) thermal expansion coefficient. Otherwise, thermal deformations in heating can cause destruction of fragile isolators. This task is to be solved with the help of sitall.

Internal polished surfaces of aluminum electrodes are covered with beryllium. Operational surfaces of quartz and sitall rings-isolators were sprayed either with beryllium oxide or compound of nickel oxide-58 and molybdenum since they are good electric isolators and have boundary rate for UCN ~ 6,8 m/s. For fixing rings and eliminating high voltage break-downs, landing ditches of 6,5 mm depth are made on the electrode-isolator boundary. Grounded electrodes are equipped with mechanism of shutters for passing neutrons and their escaping after the period of storage.

The most important characteristic of the installation is UCN storage time in the spectrometer traps. Owing to the spectrum of stored neutrons, the measured storage time is not defined by one value. Within the interval of holding neutrons in the chamber from 0 to 30 sec, the storage time is 30 – 40 sec, and in the time interval of holding them in the chamber for 30 – 120 sec, the storage time is 90 – 100 sec. Fig.7 gives curves of the number of UCNs after different periods of holding neutrons in the top and bottom storage chambers as well as for the detectors located in front of the flipper and behind it.

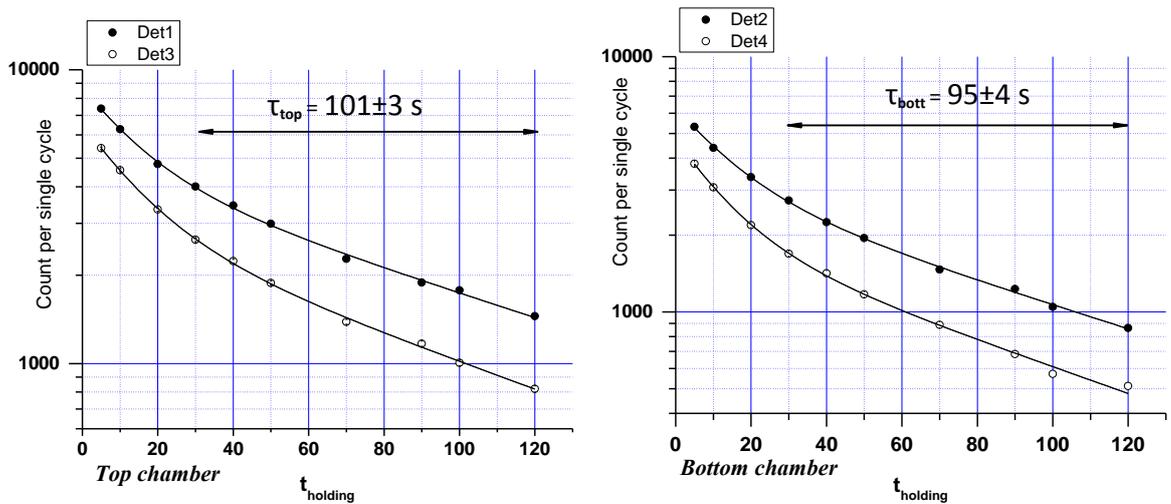

Fig.7. Storage time of UCN in traps.

### 4. High voltage source.

Compact bipolar high voltage source ±200 KV with controlled polarity in low voltage circuit was created similar to that elaborated in PNPI earlier [33, 34]. Such a source is free of weakness of mechanical switchers of electric voltage polarity. The source is connected with the spectrometer by means of 6 meter high voltage cable with polyethylene isolation, in which distilled water filling the central canal plays the role of the central leader. Distilled water is used as it has high degree of chemical purity. Resistance of water column is 200 MOhm, which combined with high internal resistance of the source considerably reduces resonance overvoltage and currents at break-downs. Isolating entrance into the vacuum chamber is made from quartz. Non-vacuum part of the high voltage quartz entrance leading into the installation via magnetic screens has an additional aluminum cover. To avoid break-downs in air gaps of the quartz entrance, the cavity between the cable and an internal surface of quartz cylinder outside the vacuum chamber is filled with insulating gas ($SF_6$). Break-downs and notable leakage currents in the cable and input do not occur until voltage rises up to ±200 kV.

A high voltage source consists of a capacitor cascade voltage multiplier placed into a hermetic vessel made from stainless steel filled with insulating gas at pressure 1.5 atmosphere. It also involves a specialized interface providing polarity commutation of high voltage at the source output without switching off the circuit of loading.

A specialized interface of the cascade voltage multiplier is related to a low voltage part of the source and is aimed at supporting control, monitoring and a low voltage power supply of the cascade multiplier. The interface is of flexible design giving possibility of operating with both manual and program control. The interface is connected with a laboratory programmed feeding source of constant current required for a power supply of cascade multiplier transformers switched subsequently to transistor keys located in the interface and forming two quasi resonance (one tact) pulse feeding sources, with each of them working for its own cascade multiplier. Resonance transformation of constant low voltage intensity into an alternating one ensures a comfortable regime of the transistor key operation as well as a break-down proof overloading characteristic of the source with current having a simple decrease of the output voltage. A program interface control and a programmed feeding source are performed by a computer via an optical communication line. In the computer there is installed a consistency plate required for modifying optical signals and a universal plate of input – output with registers (ADC/DAC) on PCI bus.

The interface does not allow either simultaneous switching of monitoring by two polarities of output intensity of a high voltage source, or forced alteration of polarity of output intensity at its non-zero value at loading.

The main units of the cascade multiplier power module are presented in Fig. 8.

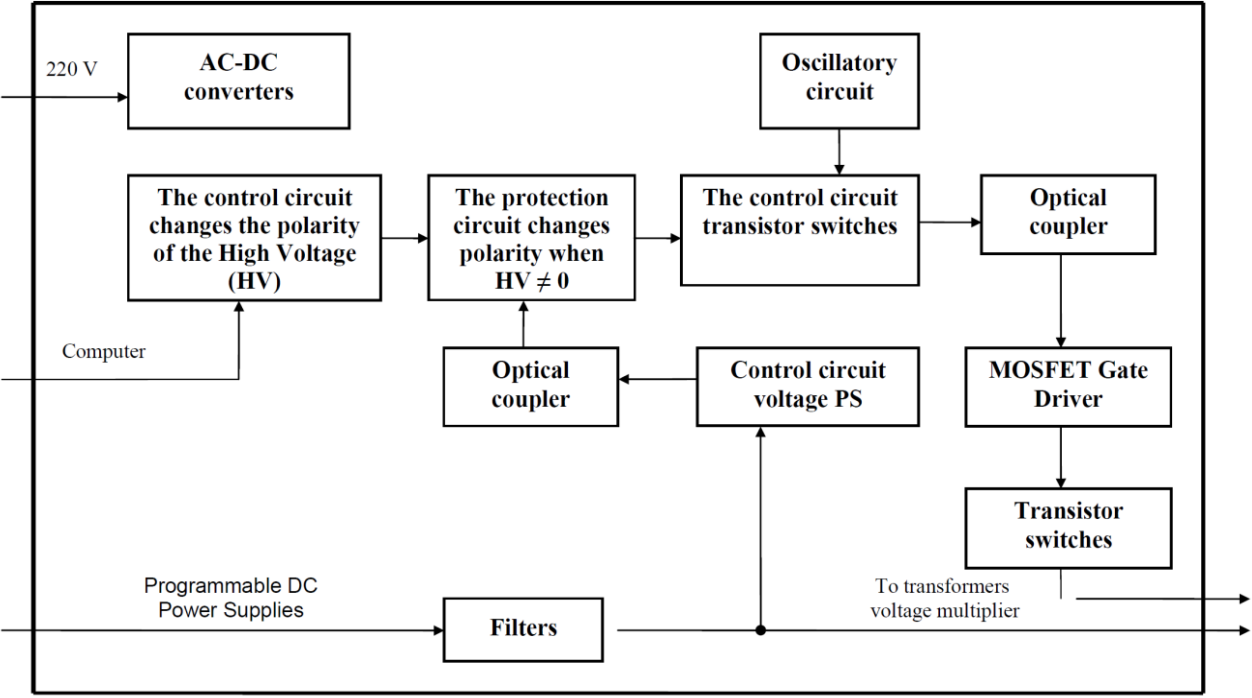

Fig. 8. Block-scheme of PMCM

**Cascade multiplier**. In fact a cascade multiplier represents two high voltage sources of different polarity switched on parallel to the total loading. At every time moment only one arm is working as high voltage source, depending on what primary transformer coil the controlled alternating current is transmitted to. Another arm operates as an additional loading, serving as voltage divider on open diodes and resistors. The latter resistances of this divider are employed for measuring the source output voltage. The voltages recorded from both resistive dividers are measured by two separate measuring channels transforming voltage into frequency. Simultaneous control of these voltages enables to monitor condition of the high voltage source itself with regard to the presence of internal leakage currents ( at normal operation of the system these voltages are equal in value but opposite in sign).

In switching polarity the cascade multipliers change their functions, which results in smooth recharging of capacitors of cascade multipliers without commutation distortions and overvoltage. As to the design, high voltage transformers and capacitor voltage multiplier of both arms of the source are assembled on ten disks-isolators located one over another. Equi-potential cascades of sources of different polarity are mounted on one disk-isolator, thus ensuring uniform distribution of potentials at the multiplier height. For smoothing out field intensity in the radial plane, every disk is surrounded with a «gradient» ring. High voltage output is closed by the screen treated with electric conduction paint, which also contributes to smoothing out electric field intensity. With the purpose of making this source more reliable, every multiplier diode cascade is made from two successively switched high voltage diodes. Fig.9 shows a simplified scheme, and Fig.10. gives an external view of the cascade multiplier.

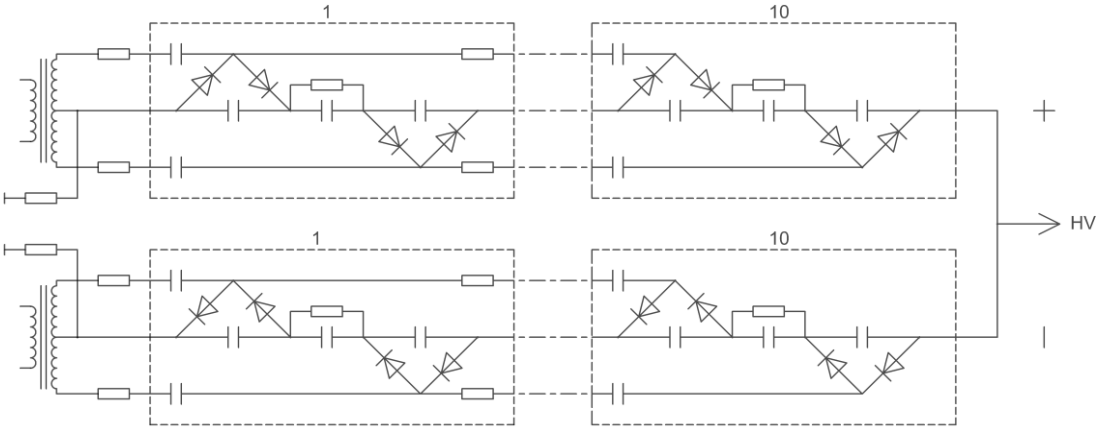

Fig. 9. The scheme of a cascade voltage multiplier.

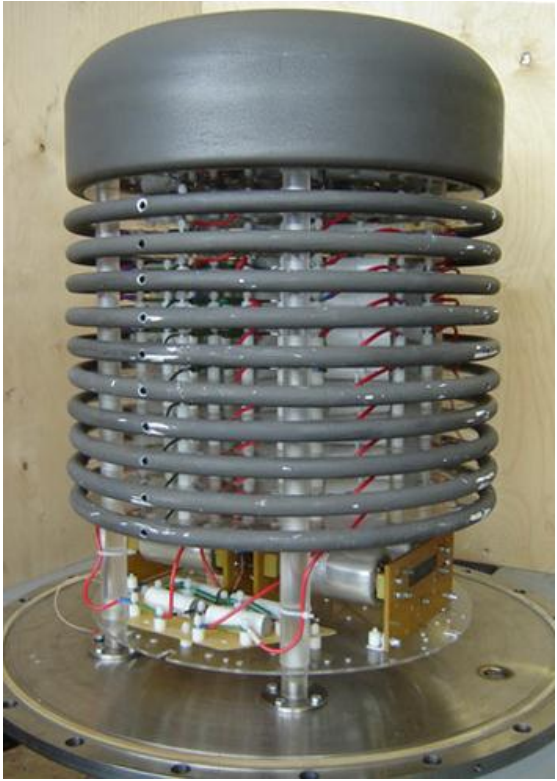

Рис. 10. Exterior view of a cascade multiplier.

Cascade multiplier is assembled in the seal shell made from stainless steel and located on a special platform; seal plugs are carried into the lower part of the cover. Use of insulating gas as a suitable working atmosphere for high voltage source enables to decrease its exterior parameters.

The output voltage value, leakage currents in loading and break-downs are registered with PC, for this purpose the voltage drops on measuring resistors of the source and loading are transformed into frequency. The data transformed into frequency are transmitted onto the universal plate of counters on PCI bus by means of an optical communication line, with measuring facilities being fed by accumulators, which provides absence of ground contours and enhances protection of the system on the whole from distortions.

PC automatically maintains the high voltage value at which leakage currents do not exceed the limit predetermined in the program. The system possesses shielding from high currents in loading: break-down currents, those of possible leakage in the source or short circuit current.

Current in loading (leakage current in UCN storage chambers) is measured in regard to voltage drop at additional resistance of 10 Ohm, switched successively between «zero» electrodes and «ground» of the high voltage source. Recording of bipolar pulse signals (caused by electric break-downs either at one or the other polarity) is carried on by one registering channel.

Constant blowing of dry pure He at pressure ~ $(1-2) \cdot 10^{-3}$ mbar via the spectrometer permitted us to diminish total leakage currents of high voltage system and increase operational voltage at electrodes in UCN storage chambers. Measurements of neutron EDM at the existing installation were performed at electric field voltage V = ±(12–14) kV/cm, while in the most accurate series of measurements made in ILL one attained the field voltage equal to 18 kV/cm. Leakage currents in chambers at various periods of time were different, depending on the condition of isolator covers and made up the value from a few tens to a few hundreds of nanoamper. In sorting out the measurement data in terms of the value of leakage currents it was shown that the attained precision level of such currents did not produce any influence on the measured value. In increasing the installation sensitivity, for example, by means of considerable enhancing of UCN intensity, one should take measures for reducing leakage currents in the UCN storage chambers.

## 5. *Magnetic field in the spectrometer.*

The four layer magnetic screen made from permalloy protected spectrometer from outer magnetic perturbations with shielding factor K ≈ 1000. For obtaining better homogeneous magnetic conditions in the operational area, magnetic screens were demagnetized after every dissembling and assembling of the spectrometer. For this purpose one transmitted current of alternating direction at frequency of (0,5–1) Hz and initial amplitude of 20A. Current magnitude was smoothly decreasing down to zero (linearly or exponentially).

Constant magnetic field $B_o = 1,8 \cdot 10^{-6}$ T is created optimally by current rings located inside magnetic screens for obtaining homogeneous magnetic field in the volume of UCN storage. The current source of these rings possesses a long term relative stability ~$10^{-5}$. Oscillating magnetic field is induced by the coil consisting of four loops around the UCN storage chambers. Thus, constant and oscillating magnetic fields are total for both chambers, while electric fields have opposite directions in the top and bottom chambers making the spectrometer scheme differential regarding the searched effect to be produced by neutron EDM.

Measurement of the magnetic field of the spectrometer is performed by 8 cesium magnetometers surrounding UCN storage chambers. They are placed in pairs in four vertical cylinder cavities fixed on the cover of the vacuum chamber (Fig.5, Fig.11 show the view from the above to the vacuum chamber cover). As far as the magnetometer design enables, the four bottom magnetometers and the four top magnetometers are located as near as possible to the bottom and top neutron storage chambers, respectively. The data obtained with these magnetometers are used by the system of stabilizing neutron magnetic resonance.

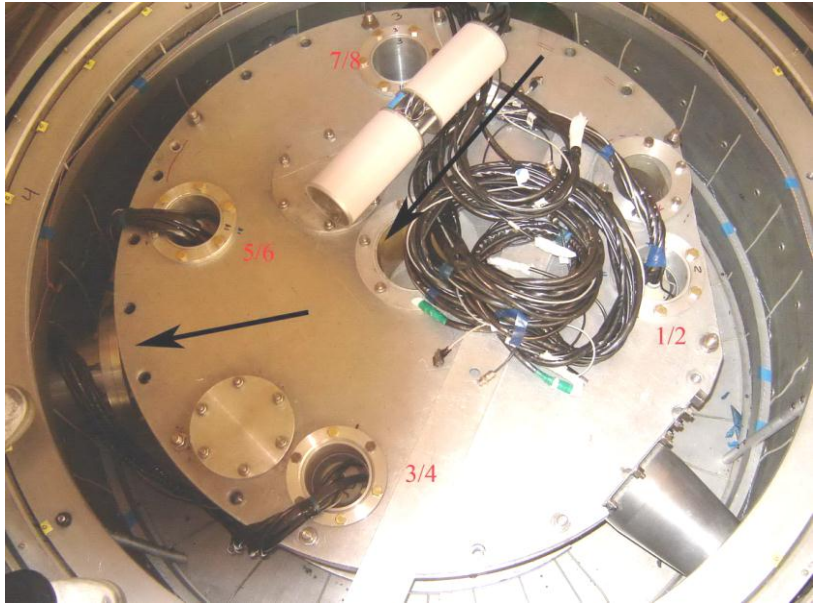

Fig.11. Magnetometric channels of the EDM spectrometer and magnetometric block from two magnetometers with light guides.

The most suitable for operation of such a system of stabilization is the self-generating magneto-meter (of $M_x$ type) based on the optical orientation of alkaline atoms [35]. This fact is essentially related to its high magneto- variation sensitivity due to the possibility of attaining a significant degree of atom polarization and small width of the magnetic resonance line by using cells with a paraffin coating, as well as the high operating speed peculiar to the $M_x$ circuit.

In accordance with it, a prototype of the cesium self-generating magnetometer for the EDM experiment [36] was designed. It was demonstrated that the calculated sensitivity of this magnetometer (in the magnetic field $B= 2\times 10^{-6}$ T) was ~$(1–2)\times 10^{-6}$ fT at a 100-s measurement time. It should be noted that in contrast to the standard circuit of the $M_x$ magnetometer, this design included a pair of irregular-type light guides, one of which was used for supplying the pumping radiation to the chamber with cesium atoms; the other was used for optical recording of signals of the Larmor precession of these atoms. However, the galvanic coupling remaining in the magnetometer circuit due to the presence of electric cable connecting the magnetic sensor to the feedback amplifier inevitably leads to failures in operation of the electronic equipment caused by electromagnetic noise (inducing pickups in wires and closed loops). This, in turn, can lead to systematic errors in long-term and accurate measurements of the neutron EDM.

In works [37, 38] one proposed the design of the noise-immune self-generating cesium magnetometer, which realizes the complete optical isolation of the magnetic sensor from the electronic equipment. Fig.12 shows a block diagram of the designed noise-immune cesium magnetometer. The magnetometer consists of the magnetic sensor placed inside the EDM spectrometer and electronic equipment located outside it. The electronic equipment includes a feedback amplifier and a pumping light source made as a high-frequency current-stabilized generator, which is intended for exciting discharges in the cesium spectral lamp. The sensor is coupled with the electronic equipment by three 5-m-long flexible multiple-strand light guides. The magnetometer operates as follows. The light from the spectral lamp excited by the high-frequency generator passes through the light guide, prism, lens, infrared filter ($\lambda_0$= 8943 nm) and circular polarizer and enters into the absorption cell shaped as a glass bulb with an inner paraffin coating and filled with cesium vapors. During the magnetic resonance, the light passing through the cell proves to be modulated with the Larmor frequency. Having passed through the lens, prism, and light guide, the light arrives at the photodetector built in the feedback amplifier board, where it is converted into the alternating current (AC) signal amplified by the wideband amplifier with a specified phase characteristic. The output voltage of the amplifier (with the Larmor frequency) controls the luminous power of the light source based on the light guide and operational amplifier that are connected in the circuit ensuring the linear dependence of the luminous power on the control voltage. The modulated light arrives at the photodiode through the light guide. The photodiode is connected to the radio frequency (RF) coil of the sensor through a capacitor with a small leakage (in order to exclude the initiation of the additional constant magnetic field). As a result, the photodiode converts the modulated light into the alternating current creating the oscillating magnetic field in the RF coil. Thus, the positive feedback loop is closed. This feedback maintains continuous oscillations in the magnetometer, the Larmor frequency of which is proportional to the measured magnetic field $f=\gamma/2\pi B$ (where $\gamma/2\pi = 3.5$ Hz/nT for cesium atoms).

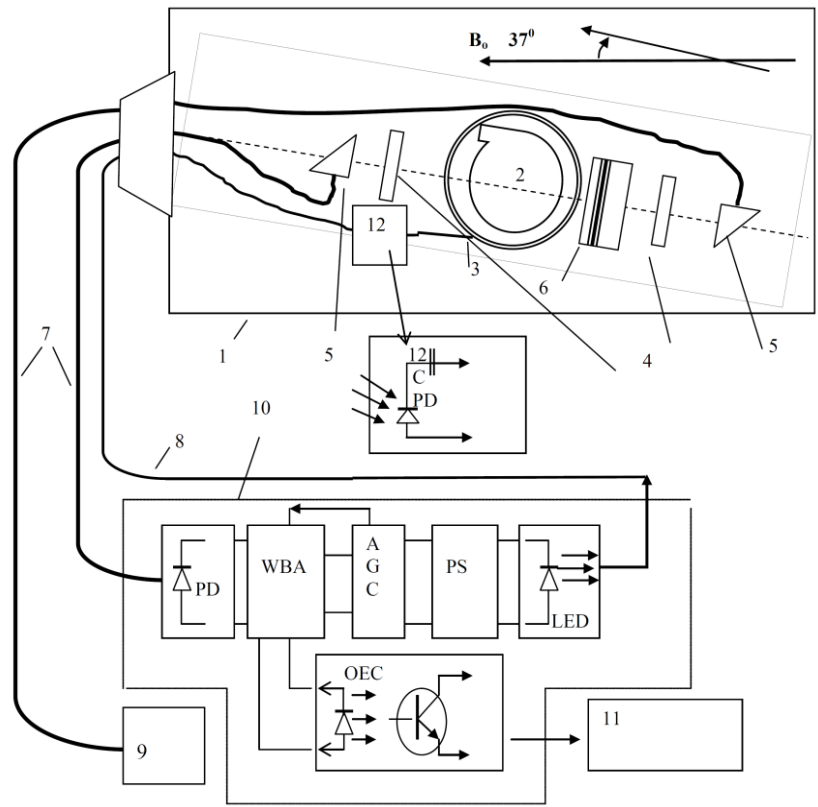

Fig. 12. Block diagram of the noise-immune cesium magnetometer: (*1*) platform made of Plexiglas; (*2*) working cell; (*3*) RF coil; (*4*) lens; (*5*) prism; (*6*) circular polarizer and $D1$ filter; (*7,8*) light guides; (*9*) spectral lamp unit; (*10*) feedback amplifier; (*PD*) photodiode, (*WBA*) is the wideband amplifier, 0.01–2 kHz; (*AGC*) automatic gain control unit; (*PS*) phase shifter; (*LED*) light-emitting diode, part of the optoelectronic device; (*OEC*) optoelectronic couple; (*11*) frequency meter; (*12*) part of the optoelectronic device placed on the magnetometer platform (*PD* is the photodiode of nonmagnetic design, and *C* is the blocking capacitor).

### 6. *Stabilization of the magnetic resonance.*

Some uncontrolled changes in magnetic conditions in UCN storage chambers lead to resonance frequency shifts, additional spread of measuring results and, hence, to diminishing the installation sensitivity. In addition to passive ways of stabilizing magnetic fields, this work used the method of «dynamic» magnetic resonance stabilization [39, 40]. If frequency of oscillating magnetic field $B_1$ at every time moment corresponds to the value of field $B_o$, this provides phase fluctuation coherence of oscillating field with precession of magnetic moment of neutron assembly. To obtain frequency proportional to the mean value of the field $B_o$, one makes use of the mean frequency of eight quantum self-generating magnetometers with optical orientation of $^{133}$Cs atoms surrounding the top and bottom neutron storage chambers within the influence of homogenous magnetic field $B_o$.

Their sensitivity to magnetic field variations is ~ 1 fT for measurement time of 100 s, as determined by the level of their intrinsic noises. These are negligible compared to the level of magnetic field perturbations inside the shield. A detailed study of the magnetometers in the magnetic screen is presented in [36]. They are placed around UCN storage chambers in pairs symmetric to the plane defined by the central electrode. The whole assembly provides monitoring of the mean magnetic field within the resonance range. As in previous measurements [23] we used the stabilization method of resonance conditions [41], with electronics upgraded for modern digital signal processing [42, 43].

Since gyro-magnetic relationships for Cs atoms and neutrons are concerned with ratio $\gamma_{Cs} \approx 120 \cdot \gamma_n$, one gets resonance frequency of oscillating field $B_1$ by dividing precession frequency of $^{133}$Cs atoms in the field $B_o$ by coefficient k=120. High rapid operation of self-generating quantum magnetometers provides synchronization of frequency and phases of field $B_1$ with alterations of field $B_o$.

The present experiment for producing RF of the field at neutron magnetic resonance frequency employs a digital synthesizer receiving signals from 8 cesium magnetometers surrounding UCN storage chambers. It is to be noted that at first one fulfills the analog processing of signals from 4 magnetometers placed in one plane (around the top and bottom neutron storage chambers), in order to obtain averaged Larmor frequency of cesium atoms. The principle of analog processing is simple: two input signals are successively passing via band pass filter, further these two analogue sinusoidal signals achieve to the four quadrature analogue multiplier. From a multiplier output the resulting signal consistently passes through the filter of top frequencies, thus suppression of entrance signals with frequencies F1 and F2 and allocation of an analogue total signal with (F1+F2) frequency is carried out. Analog signal at (F1+F2) frequency achieves the comparator block in which it is converted into logical signal of meander kind (analog- to- logic conversion occurs at the intersection of analog signal zero). Further, divided by 2 on frequency a logic signal arrives on the band pass filter in which the basic harmonic of a logic signal with frequency (F1+F2)/2 or averaged frequency for two magnetometers is allocated. The averaged frequency for magneto-metric quartet is produced according to this scheme block in Fig.13.

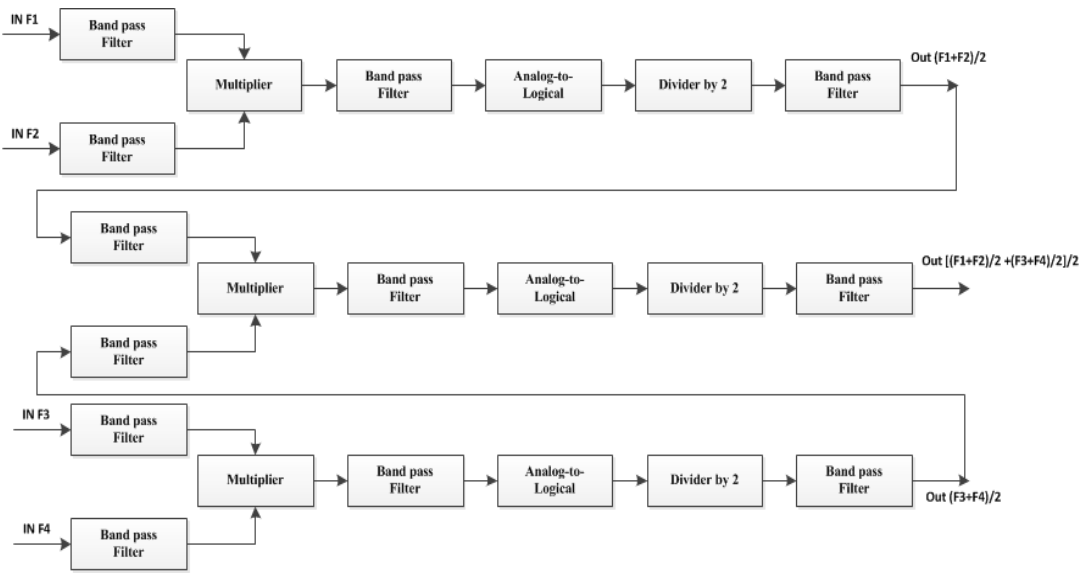

Fig.13. The block scheme for averaging of frequency for quartet of magnetometers.

In a similar way there occurs an averaged frequency of two magneto-metric quartets (Fig.14). Obtained in this way the signal with averaged frequency of magnetic resonance for cesium atoms arrives on a digital synthesizer. The digital synthesizer is aimed at producing frequency (~ 58 Hz) of neutron magnetic resonance by means of dividing frequency (~ 6957 Hz) of averaged cesium signal by coefficient K=119.95959. Division of analogue frequency is carried out by means of integrated microcircuit DDS (Direct Digital Synthesis). In our case the 32-bit synthesizer provides the variations of the dividing coefficient with step $3.3 \cdot 10^{-6}$.

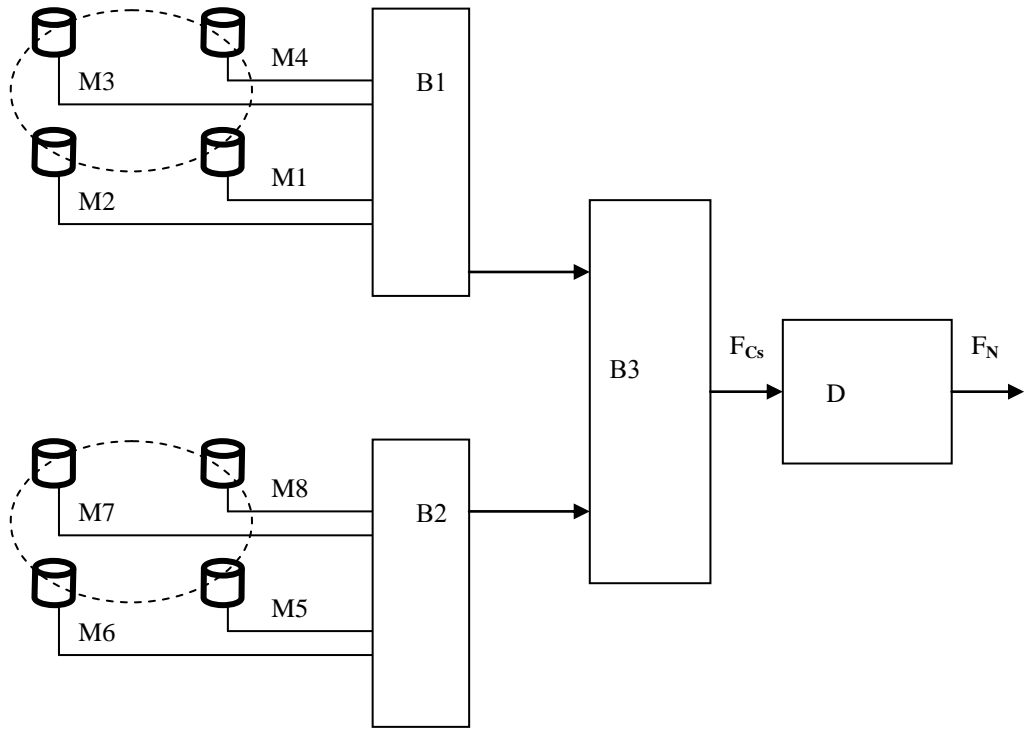

Fig. 14. System of producing radio field with frequency of neutron magnetic resonance: M1–M8 – cesium self-generating magnetometers, B1 and B2 – devices for producing averaged cesium frequency for top and bottom neutron chambers, correspondingly, B3 – block for inducing average frequency for two chambers, D – digital synthesizer of the radio field signal with frequency of neutron magnetic resonance.

Magnetic field behavior in the spectrometer depends on external magnetic perturbations and intrinsic noise. Fig.15 presents two cases of magnetic field behavior: 1) quiet magnetic conditions in which measurements were usually made; 2) conditions of large magnetic perturbations, which arose when a bulky bridge crane was working in the vicinity of the installation. Measurements were not carried on at this period of time. It is to be noted that half width of neutron resonance at the time of UCN holding in the spectrometer trap for 100 s is equal to $5 \cdot 10^{-3}$ Hz, which corresponds to magnetic field alteration 0.17 nT. In quiet magnetic conditions magnetic field fluctuations are available within resonance half-width, which appears to be not quite sufficient for the spectrometer reliable operation, hence, application of the dynamic stabilization system of resonance conditions is of great necessity. Superposition of resonances in top and bottom chambers was achieved by current in the spectrometer gradient coils.

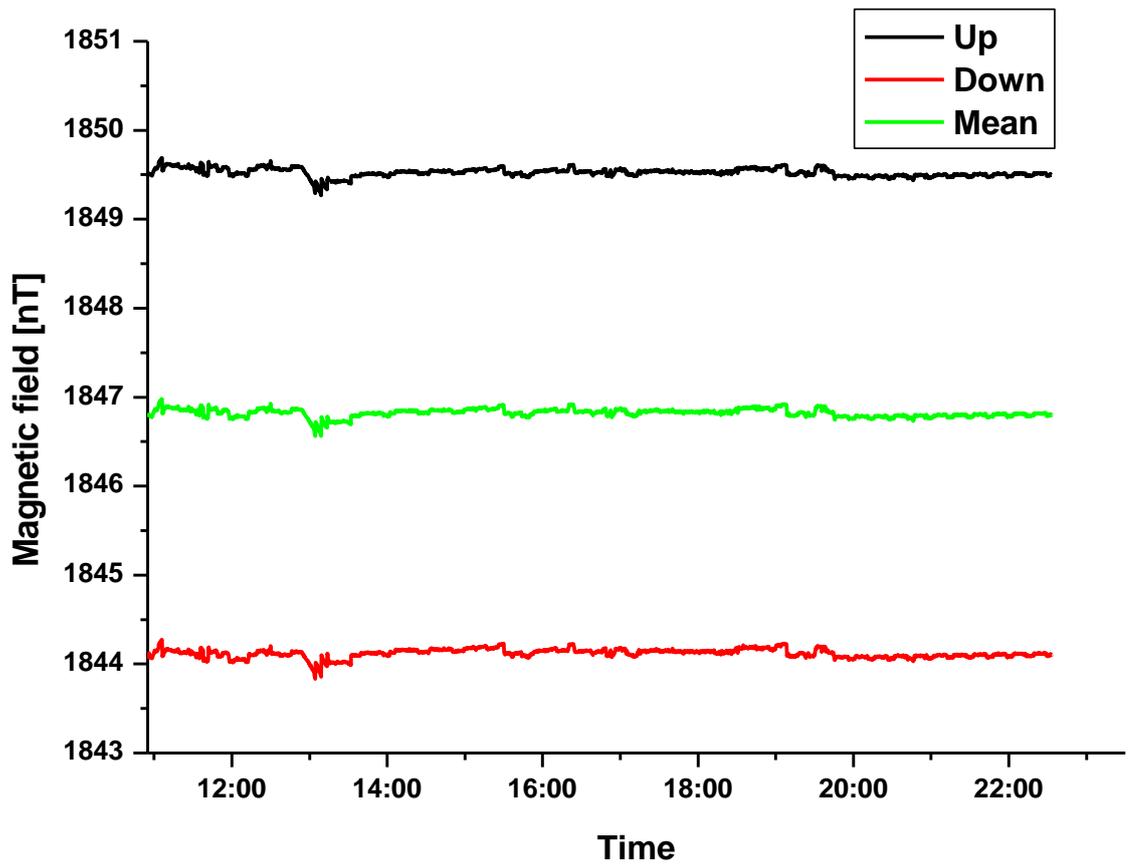

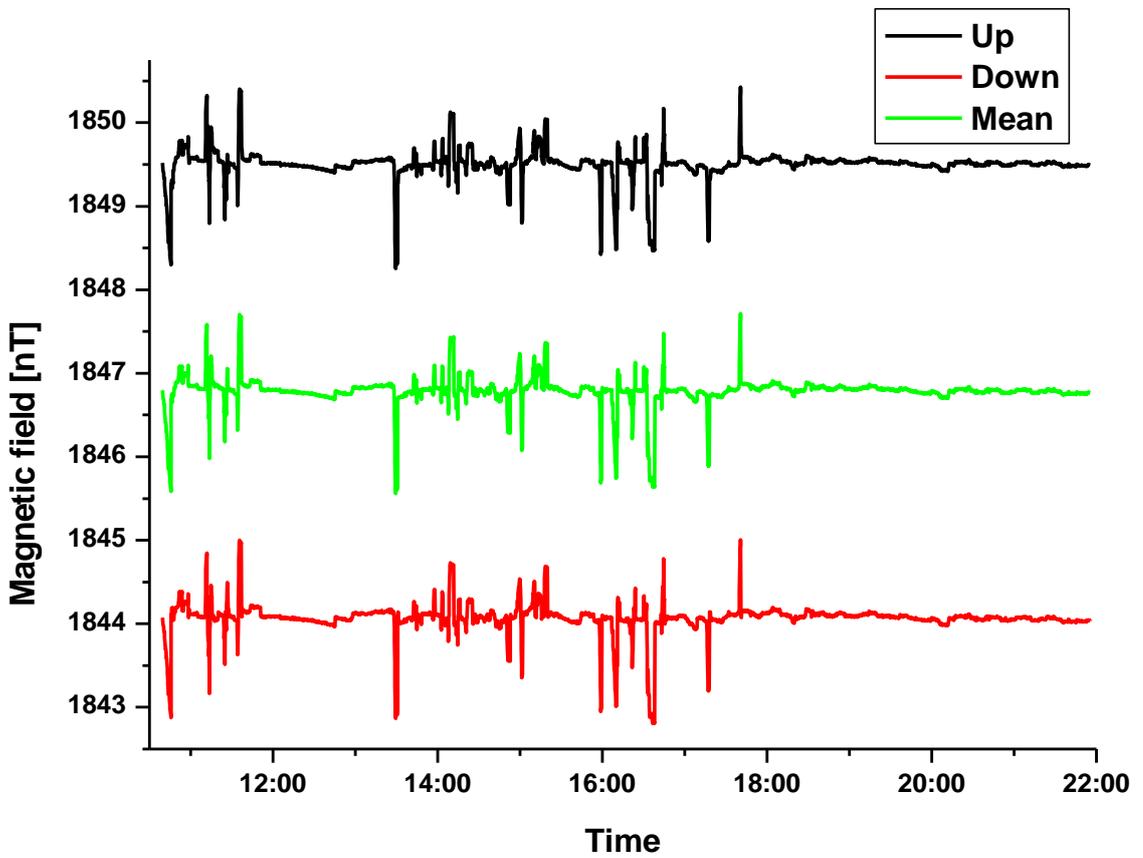

Fig. 15. Behaviour of the magnetic field in the spectrometer.

а) smooth magnetic situation. Here are presented values of the magnetic field in the upper and lower plates of the spectrometer (averaged on four), as well as average value of the magnetic field on 8 magnetometers.

б) testing of work for the whole day of the bridge crane operation after prophylactic in an unquiet magnetic situation. Correlated changes of the magnetic field with amplitude up to 1nT are observed.

In order to investigate the stabilization system of resonance conditions, a specified experiment was conducted, in which instead of UCN storage chambers an additional magnetometer was installed in the spectrometer center. It measured magnetic field values, while 8 magnetometers located around determined an average value of magnetic field in peripheries.

Fig.16 shows how well magnetic field in UCN storage chambers is described by the mean value determined in terms of recordings of eight Cs-magnetometers surrounding the operational volume. Measurements are made in the conditions of relatively small outside magnetic perturbations caused by the bridge crane operation at a large distance from the installation. The upper graph shows average frequency of eight Cs-magnetometers (divided by k=120). The middle graph gives the correspondent frequency from an additional magnetometer. Difference in values is shown in the lower graph. As frequency of field $B_1$ is obtained from average frequency value of 8 Cs-magnetometers, the suppression factor of influence of external magnetic fluctuations on resonance condition stability in EDM spectrometer («dynamic» stabilization factor) is 10 – 15 times.

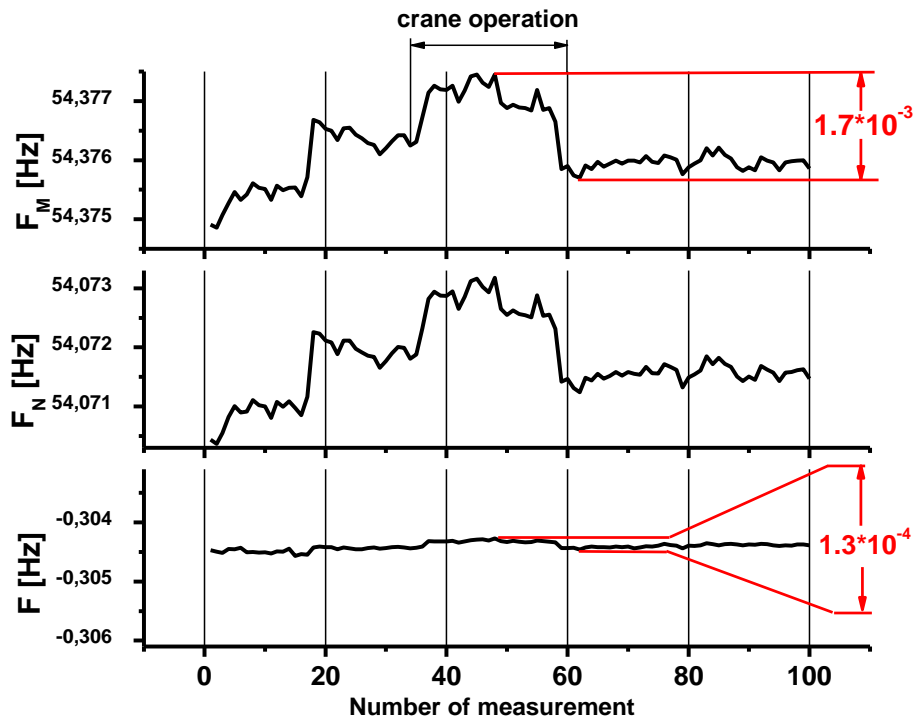

Fig.16. describes the average field quality in UCN working volume with the help of eight Cs-magnetometers.

For receiving resonance curves and experimental measurement of derivative in the working point, one applies the method of resonance tuning out by means of frequency alteration of radio pulses controlling neutron polarization at the beginning and the end of their storage. Frequency scanning was carried on by a special electronic scheme with controlled fraction coefficient of division. The scheme produced the required frequency of oscillating field from average frequency of eight magnetometers as described above. Division coefficient was changed by the control program within the predetermined range approximately at the value of k=120 with the required pitch and relative accuracy not lower than $10^{-5}$.

For compensation of slow changes of the magnetic field gradient and superposition of magnetic resonances in the top and bottom storage chambers, the measurement program of neutron EDM was focused on correction of the working point (division coefficient) in terms of alteration in neutron detector counts and in value and sign of the measured derivative. At resonance discrepancy of more than ¼ period, one could make correction of the magnetic field gradient employing current of special coil for resonance superposition. Application of such a technique could diminish spread of measuring results and reduce final experimental errors practically to the level of statistical spread.

***Diagram of measurements.*** Typical cycle of measurements was as follows: filling storage chambers with neutrons – 50 s, UCN storage in the resonance area at constant electric field at definite polarity – 70-100 s, the neutron output and detection with polarization analysis – 50 s. To provide resonance of Ramsey type at the beginning and the end of storage period, two radio pulses were applied at resonance frequency of about 3 s and phase shift between them ±90°.

High voltage polarity was switched in the interval 50 s between measuring cycles in the following sequence $+--+$ or $-++-$. Duration of such blocks from four measuring cycles was about 15 min.

### *7. Sources of random and systematic errors of measurements*

Magnetic field perturbations in UCN storage chambers alter resonance frequency and result in changing the neutron count in detectors. This is responsible for an additional spread of results as compared with Poisson statistics. External magnetic noises (magnetic condition changes in the experimental hall) are suppressed to a certain extent with passive magnetic shielding (four layer magnetic screen from permalloy) and with the help of described above technique of «dynamic» stabilization.

Internal sources of magnetic noise are likely to be parasite currents on metallic cover of the vacuum chamber, on internal permalloy screen and on other conducting units of the installation. For excluding possibility of emergence of large closed current contours, all big parts of the installation are isolated from each other and grounded at one common point. Another source of distortions is magnetic field caused by discharges and recharging currents at switching electric field in the high voltage system contour. Magnetic field pitches induced by these currents can be recorded owing to hysteresis of magnetic screens. At the installation sensitivity gain one should take into account magnetic noise made by fluctuations of thermal currents of free electrons in electrode material. Since electrodes are used as walls in UCN storage chambers, such an effect can be observable.

Non-incidental alterations of magnetic condition in resonance volume correlated with electric field switching result in systematic false effects. They can either cause distortions or imitate the searched effect of neutron EDM. Correlation data analysis of the multi-count detecting system, leakage currents and magnetometer data recordings, in principle, allows revealing some of them "a posteriori".

As mentioned above, proportional to neutron velocity and imitating EDM the false effect because of lack of parallelism of **E** and **B**$_o$ (effect "[**v**×**E**]"), restricting possibilities of beam experiments on thermal and cold neutrons, must disappear in employing a spectrometer with UCN storage due to the lack of distinctive direction of neutron movement under electric field influence.

### *8. Sensitivity of the installation and results of measurements.*

From the counts of each of the detectors D1–D4 one can derive the corresponding experimental values $d_1$-$d_4$ for the neutron EDM. Compensation, respectively, determination of some systematic effects is obtained by using linear combinations of the individual values $d_i$:

$$EDM = \frac{1}{4}[(d_1 + d_2) + (d_3 + d_4)]$$

$$\Delta \nu = \frac{1}{4}[(d_1 - d_2) + (d_3 - d_4)]$$

$$\Delta N = \frac{1}{4}[(d_1 - d_2) - (d_3 - d_4)] \quad (3)$$

$$Z = \frac{1}{4}[(d_1 + d_2) - (d_3 + d_4)]$$

The first equation determines a value for the neutron EDM. For a fully symmetric setup, this combination of values $d_i$ completely compensates fluctuations common to both chambers and correlated with reversal of the electric fields. Moreover, the linearly independent second and third combinations in eq. (3) give measurements of such fluctuations. While $\Delta \nu$ defines the effect of electric field influence on resonance conditions, $\Delta N$ measures a systematic effect on neutron count rates. Finally, in the last combination $Z$ all aforementioned effects (including the neutron EDM) are compensated, so that the condition $Z = 0$ provides a crucial test of the compensation scheme.

Thus, the cited scheme allows monitoring possible systematic errors in the course of measurements. Moreover, this scheme is capable of compensating effect $\Delta \nu$, in case of homogeneous changes of magnetic field correlated with high voltage switching. The same is true concerning compensation possibility of effect $\Delta N$, in case of similar influence on all count tracts. A considerable degree of compensation is not to be expected. One should take into consideration appearance of statistically meaningful effects (over the third – the fourth deviations), find out their reasons and take measures for their elimination. As shown above, compensation degree of homogenous magnetic perturbations makes up approximately 10 -15 times (see Fig.16).

In order to perform measurements of values $d_i$ one first has to determine the working point where the frequencies of free neutron precession and the rf field are exactly the same. This is done with neutron measurements for different time intervals $T$ between rf pulses, the working point being independent of them. Highest sensitivity to the neutron EDM is achieved if one applies phase shifts $\Delta \varphi$ of 90° or 270° between the two rf pulses. As shown in Fig. 2 this leads to the largest slope $\partial N/\partial f$ of neutron counts at the working point (compare with eq. (2)). For each of the two phase shift settings this slope is determined from two neutron count rates obtained with small shifts of the radio-frequency by about $\Delta f = \pm 5 \times 10^{-4}$ Hz around the working point. The fact that the signs of the slopes are different for $\Delta \varphi = 90°$ and $\Delta \varphi = 270°$ is very useful to monitor drifts of the resonance curve used to provide additional stabilization of resonance conditions (note that for detecting influences of electric field polarity alteration on the neutron count rate (i.e. $\Delta N$ in eq. (3)) measured rates for $\Delta \varphi = 90°$ and $\Delta \varphi = 270°$ must be subtracted rather than added to account for the sign of the slope).

Measurements of the neutron EDM are performed within a sequence of different settings of electric field polarity, $\Delta \varphi$ and $\Delta f$. This sequence is chosen with the goal to minimize unwanted effects of drifts of the resonance condition. For instance, with $\Delta \varphi$ and $\Delta f$ in a given state the electric field polarity is alternated according to the sequence (+ − − +) or (− + + −), which eliminates the effect of linear drifts if measurements are performed within constant time intervals.

The measurements were carried out at the ultracold neutron facility PF2 at the ILL in Grenoble, France. As determined in prior experiments the beam port PF2/MAM provides a UCN number density of about 7.5 n/cm$^3$ at the entrance of the EDM spectrometer [44]. Figure 17 presents an exemplary series of experimental data taken during 15 hours, with an electric field of 18 kV/cm and UCN storage time of 100 sec. Each point represents the result of a single measurement sequence of the value of EDM in accordance with eqs. (1) and (3). Also quoted are separate results on the quantities $EDM_{top} = (d_1+d_3)/2$ and $EDM_{bott} = (d_2+d_4)/2$ for the top and bottom chambers. Total results from the series shown in Fig. 5 are $EDM_{top} = (2.59\pm3.90)\times10^{-25}$ e·cm, $EDM_{bott} = -(3.98\pm4.22)\times10^{-25}$ e·cm and $EDM = -(0.70\pm2.17)\times10^{-25}$ e·cm. The latter permits us to assess the sensitivity of the experiment when running under smooth conditions, which for this series of measurements amounted to $1.7\times10^{-25}$ e·cm/day.

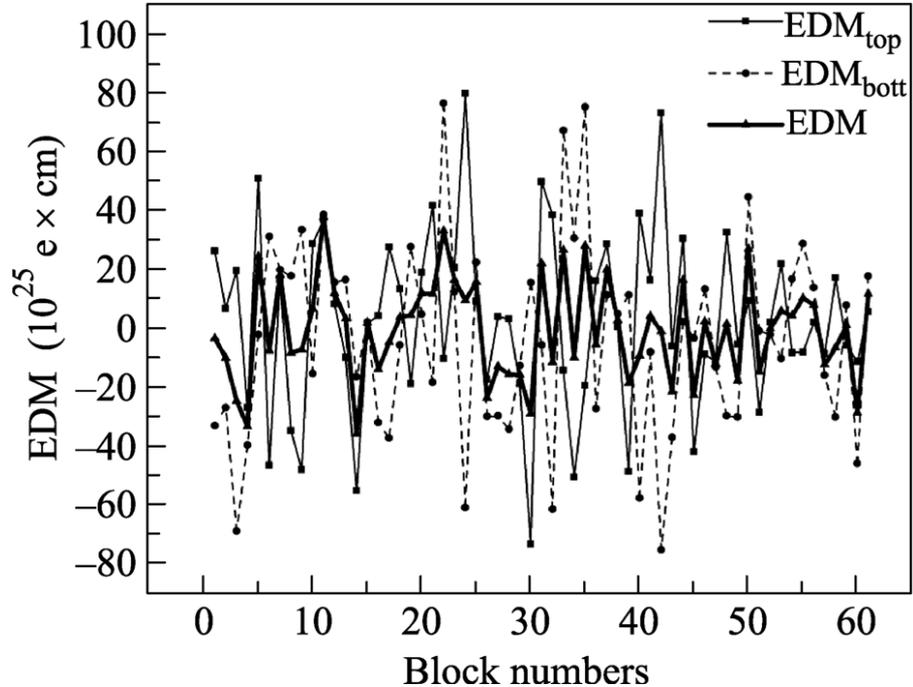

Fig.17: Exemplary series of measurements. $EDM_{top}$ and $EDM_{bott}$ are values measured for the top and bottom chamber separately (see text). $EDM$ is the quantity defined in eq. (3) that measures the neutron EDM using measurements with both chambers and all 4 UCN detectors

Several of such series as shown in Fig. 3 were obtained at the ILL during three reactor cycles of 50 days each. Since after these measurements the reactor has been shut down for an extensive period of ten months for maintenance and upgrades, we felt it timely to present our first results. Results for EDM are quoted in Table together with measured values of the quantities $\Delta v$, $\Delta N$, and $Z$ defined in Eq. (3), all in units of $10^{-26}$ e·cm (these results were previously published in[45]). The deviations from zero of the values EDM and $Z$ do not exceed one standard deviation. The values of $\Delta v$ and $\Delta N$, on the other hand, deviate from zero by two and 3.5 standard deviations, the latter with different signs in the two measurements.

Table 1. Results of measurements in units of $10^{-26}$ e·cm.

|  | Previous (PNPI) [23] | New (ILL) [45] | Total |
|---|---|---|---|
| *EDM* | 0.7±4.0 | 0.36±4.68 | 0.56±3.04 |
| $\Delta\nu$ | -22.8±9.2 | -10.04±5.98 | -13.8±5.01 |
| *ΔN* | -14.5±4.4 | 18.62±5.15 | -0.53±3.35 |
| *Z* | -0.8±4.0 | 3.68±4.72 | 1.05±3.05 |

Since the crucial criterion $Z = 0$ is fulfilled, we assume systematic effects to be visible in Δν and ΔN as compensated in the value EDM. As mentioned above, our spectrometer scheme can compensate homogenous changes, while inhomogeneous magnetic field alterations resulting from leakage currents cannot be fully compensated. Therefore, special attention must be paid to an experimental data analysis depending on leakage currents.

In particular, local alterations of the magnetic field associated with leakage currents may escape detection by an assembly of a few discrete cesium magnetometers. In the reported measurements the leakage current during a typical data set was a few tens of nanoampere and it did never exceed 2000 nA.

To test possible influence of leakage currents on measurement result of EDM and quantities Δν, ΔN, Z, series were grouped on the basis of mean value of leakage currents measured during neutron storage. Current intervals of groups were chosen as follows: 0-50 nA; 50-100 nA; 100-150 nA; 150-300 nA; 300-500 nA; 500-1000 nA and 1000-2000 nA. Fig.18 presents these results. $\chi^2$ characterises spread of points for a hypothesis of a zero average.

As non-identical changes of magnetic field in the top and bottom neutron storage chambers are not compensated with a differential spectrometer, the lower graph shows dependence on leakage currents of magnetic field gradient values in the spectrometer, measured in the same time intervals. Gradient is calculated as difference between average values of upper and lower magnetometers during the neutron storage time. Current leakage dependence is not observed in either of the presented graphs.

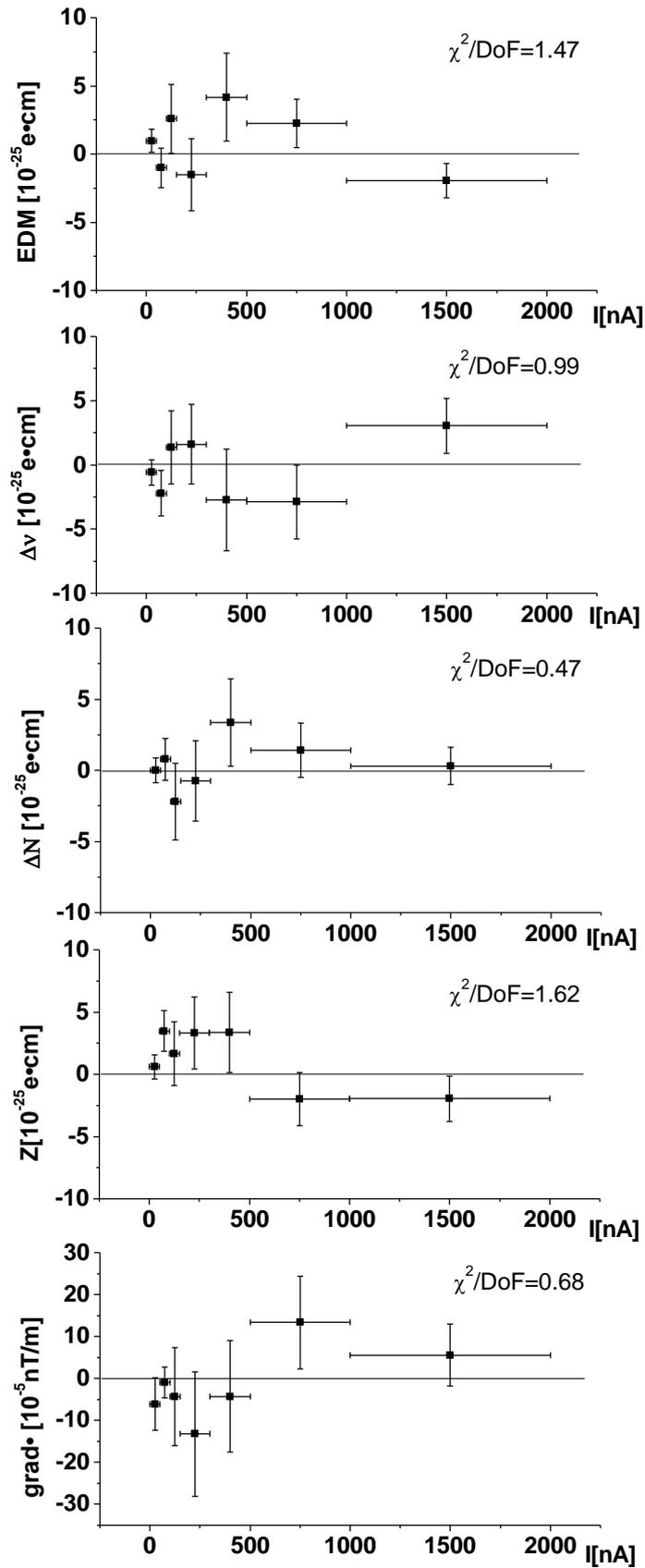

Fig. 18. Results of EDM measurements for different intervals of leakage current.

A separate control of possible effects on the cesium magnetometers has shown that after switching the electric field the difference in recordings of the magnetometers in the upper and lower planes (see Fig.5) did not exceed 1 fT determined with all data accumulated during the EDM measurements. This limit can be converted into a limit of a corresponding false EDM effect of $2\times 10^{-27}$ e·cm, i.e., by still more than a factor ten below the quoted accuracy of measurements.

Distributions of measured normalized values of EDM, $\Delta\nu$, $\Delta N$, and $Z$ are presented in Fig. 19 (normalized values are $Y_i = (y_i - <y>)/\sigma$, where $y_i$ are measured values, $<y>$ is mean value, $\sigma$ is the standard deviation).

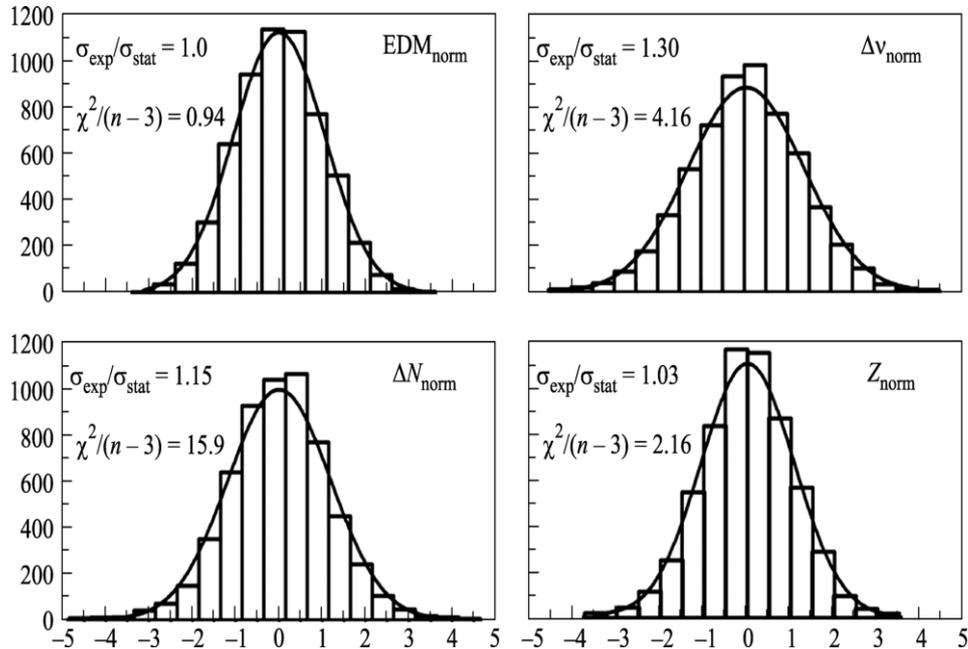

Fig. 19. Distribution of measured values for the quantities defined in Eq. (3). For each quantity is shown the ratio of the width of the measured distribution to its uncertainty defined by counting statistics, as well as $\chi^2$ obtained from fitting a normal distribution.

The width of the distributions of values EDM and $Z$ corresponds to that of a normal distribution defined by counting statistics. The distributions of values of $\Delta\nu$ and $\Delta N$ are somewhat broadened due to imperfect stability of magnetic field and neutron intensity. The absence of any broadening of the EDM and $Z$ distributions demonstrates the compensation of fluctuations of the magnetic field.

The accuracy of the reported measurements of neutron EDM at ILL was $4.7\times 10^{-26}$ e cm. Including the earlier result obtained at the WWR-M reactor at PNPI [23] the total accuracy is $3.0\times 10^{-26}$ e cm. We interpret our new result as a limit on the neutron electric dipole moment of $|d_n| < 5.5\times 10^{-26}$ e·cm at 90% confidence level. This experiment involves different systematic effects than other neutron EDM experiments. Notably it provides a result free from false effects due to geometric phases.

For upcoming runs of the experiment we plan to relocate the apparatus to another ILL UCN beam which should bring a 3 to 4 fold gain in intensity at the entrance of the spectrometer. In addition, some redesign of the neutron transport within the spectrometer should provide a further increase in neutron count rates. With these measures in place the accuracy might be improved in near future down to $10^{-26}$ e·cm. Progress beyond that level will require novel UCN sources such as the upgrade of the source described in [46] projected at the ILL, or in a later stage a new source at the reactor PIK with prior implementation of a prototype at the WWR-M reactor at PNPI [30].

Creation of such UCN sources opens great possibilities for increasing EDM experiment sensitivity, however, in this case requirements to the experimental facility will be much more considerable: decrease of magnetic noise in UCN storage chambers, stability of magneto-resonance conditions, control of systematic effects and their suppression. A separate task is to design UCN detectors capable of operating with high input loading. The installation sensitivity gain can be achieved by upgrading the following parameters: UCN storage time, neutron polarization as well as electric field intensity gain in UCN storage chambers. Some issues of upgrading PNPI double chamber spectrometer aimed at increasing its sensitivity have been considered in work [32].

We thank the members of the subcommittee of college 3 for continuous support and their recommendation to perform neutron EDM measurements over an extended period of 2–3 years. We are also grateful to the staff of the workshop of experimental facilities at PNPI for their assistance in reconstructing the spectrometer, and to the personnel of the ILL reactor for help in assembling the installation and doing measurements. We thank T. Brenner from ILL and A.I. Egorov, E.P. Volkov, E.V. Siber, A.A. Sumbatian, and T.V. Savelieva from PNPI for their assistance. We also acknowledge a large contribution of I. Tokmakov, N. Mikulinas, and A. Pikalev on construction and operation of electronics and high-voltage source.

Various parts of the carried out researches are executed at PNPI by support of Russian Foundation for Basic Research (project no. 14-22-00051) and Russian Scientific Foundation (project no. 14-22-00105) in full conformity with independent plans of these projects.

**References**
[1] Purcell E.M., Ramsey N.F., Phys. Rev., 1950, v.78, p.807.

[2] Lee T.D., Yang C.N., Phys. Rev., 1956, v.104, p.254.

[3] Wu C.S., Ambler E., Hayward R.W., Hoppes D.D., Hudson R.P., Phys. Rev., 1957, v.105, p.1413.

[4] Christenson J.H., Cronin J.W., Fitch V.L., Turlay L., Phys. Rev. Lett., 1964, v.13, p.138.

[5] Barr S.M., Int. Journal of Mod. Phys. A, 1993, v.8, No.2, p.209.

[6] I.Bigi, N.G. Ural'tsev, Zh. Eksp. Teor. Fiz. 100, 1991, 363-385

[7] J. Baron, W.C. Campbell, D. DeMille et al., Science, vol. 343 no. 6168 pp. 269-272 (2014)

[8] NSAC Long Range Plan: The Frontiers of Nuclear Science, 2007; [http://science.energy.gov/~/media/np/nsac/pdf/docs/2007_lrpwp_symmetries_neutrinos.pdf]

[9] M. Pospelov, A. Ritz, Ann. Phys. **318**, 119 (2005).

[10] Smith J.H., Purcell E.M., Ramsey N.F., Phys. Rev., 1957, v.108, p.120.

[11] Dress W.B., Miller P.D., Pendlebury J.M., Perrin P., Ramsey N.F., Phys. Rev., 1977, v.D15, p.9.

[12] Shull C., Nathans R., Phys. Rev. Lett., 1967, v.19, p.38.

[13] V.L. Alekseev, V.V. Voronin, E.G. Lapin et al., Zh. Eksp. Teor. Fiz. 96, 1989, 1921-1926.

[14] V.V. Fedorov, V.V. Voronin and E.G. Lapin, J. Phys. G, Nucl. Part. Phys. 18 (1992) 1133-1148.

[15] V.V. Fedorov, M. Jentschel, I.A. Kuznetsov et al., Phys. Lett. B, 2010, v.694, p.22-25.

[16] Zeldovich Y.B., JETPh, 1959, т.36, с.1952.

[17] Shapiro F.L., UPhN, 1968, т.95, с.145.

[18] Altarev I.S. et al., Nuclear Physics A341 (1980) 269-283,

[19] Altarev I.S. et al., Physics Letters B102 (1981) p.269

[20] Pendlebury J.M.et. al., Physics Letters B136 (1984) p.327

[21] Smith K.F. et al., Phys. Lett. B, 1990, v.234, p.191.

[22] Altarev I.S. et al., Physics Letters B82 (1992) p.242

[23] Altarev I.S., Borisov Yu.V., Borovikova N.V. et al., Phys. of At. Nucl., v.59, No.7, p.1152.

[24] Harris P.G. et al., Phys. Lett. B, 1999, v.82, p.904.

[25] C.A. Baker, D.D. Doyle, P. Geltenbort et al., Phys. Rev. Lett., 2006, v.97, p.131801.

[26] J. Pendlebury et al., Phys. Rev. A 70, 032102 (2004).

[27] Steyerl A., Nagel H. et al., Phys. Lett. B, 1986, v.A116, p.347.

[28] Altarev I.S., Borovikova N.V. et. al. Pis'ma Zh. Exper. Teor. Fiz., 1986, v.44, p.269.

[39] A.P. Serebrov, A.K. Fomin, M.S. Onegin et al., Technical Physics Letters, v.40 (2014) 10

[30] A. P. Serebrov, V. A. Mityuklyaev, A. A. Zakharov, et al., Nucl. Instrum. Methods Phys. Res. A 611, 276 (2009).

[31] Ramsey N.F., Molecular Beams, Oxford University Press, 1956, 1983.

[32] Altarev I.S., Borisov Yu.V., Ivanov S.N. et al., Preprint of PNPI, Russ. Acad. Sci., Gatchina, 2003, №2514, c.41.

[33] Borisov Yu.V., USSR Inventor's Certificate no. 944021 (1980). Byull. Izobret., 1982, no. 26.

[34] Borisov Yu.V. et al., Prib. Tekh. Eksp., 1987, no. 2, p.120.

[35] Pomerantsev, N.M., Ryzhkov, V.M., and Skrotskii, G.V., Fizicheskie osnovy kvantovoi magnitometrii (Physical Bases of Quantum Magnetometry), Moscow: Nauka, 1972.

[36] E.B. Aleksandrov, M.V. Balabas, S.P. Dmitriev et al., Tech. Phys. Lett. 32, 627 (2006).

[37] Borisov, Yu.V. and Slyusar' V.N., USSR Inventor's Certificate no. 1691804, Byull. Izobret., 1991, no. 42, p. 15.

[38] E.B. Aleksandrov, M.V. Balabas, Yu.V. Borisov et al., Instruments and Experimental Techniques, 2007, Vol. 50, No. 1, pp. 91–94.

[39] Borisov, Yu.V. and Ivanov S.N., USSR Inventor's Certificate no. 919477, Byull. Izobret., 1983, no. 30.

[40] Altarev I.S. et al., Preprint of LNPI, Russ. Acad. Sci., Leningrad, 1985, no. 1117, p.14.

[41] Yu.V. Borisov, S.N. Ivanov, V.M. Lobashev et al., Nucl. Instr. Meth. A **357**, 115 (1995).

[42] V.A. Solovei, V.V. Marchenkov, I.A. Krasnoshekova et al., Preprint PNPI 2565 (2004).

[43] E.B. Aleksandrov, M.V. Balabas, Yu.V. Borisov et al., Tech. Phys. Lett. **33**, 3 (2007).

[44] A.P. Serebrov, P. Geltenbort, I.V. Shoka et al., Nucl. Instr. Meth. A **611**, 263 (2009).

[45] A.P. Serebrov, E.A. Kolomenskiy, A.N.Pirozhkov et al., JETP Letters, **99**, pp. 4–8, (2014).

[46] O. Zimmer, F.M. Piegsa, S. Ivanov, Phys. Rev. Lett. **107**, 134801 (2011).